\documentclass[intlimits,twoside,a4paper]{article}

\usepackage{amsmath,amssymb}
\usepackage{graphicx}

\usepackage[T2A]{fontenc}
\usepackage[cp1251]{inputenc}

\usepackage[eqsecnum]{cmpj2}

\usepackage{psfrag}

\issue{2014}{17}{3}{33604}
\doinumber{10.5488/CMP.17.33604}

\title[Polymers in disordered environments]%
{Polymers in disordered environments%
}

\author[V. Blavatska, N. Fricke, W. Janke]{V. Blavatska\refaddr{label1},
N. Fricke\refaddr{label2}, W. Janke\refaddr{label2,label3}
}
\addresses{
\addr{label1} Institute for Condensed Matter Physics of the National Academy
of Sciences of Ukraine, 79011 Lviv, Ukraine
\addr{label2} Institut f\"ur Theoretische Physik, Universit\"at Leipzig,
Postfach 100 920, 04009 Leipzig, Germany
\addr{label3} Centre for Theoretical Sciences (NTZ), Universit\"at Leipzig,
Postfach 100 920, 04009 Leipzig, Germany
}

\date{Received June 18, 2014, in final form July 15, 2014}
\authorcopyright{V. Blavatska, N. Fricke, W. Janke, 2014}

\begin{document}

\maketitle

\begin{abstract}
A brief review of our recent studies aiming at a better understanding of
the scaling behaviour of polymers in disordered environments is given.
The main emphasis is on a simple generic model where the polymers are
represented by (interacting) self-avoiding walks and the disordered
environment by critical percolation clusters. The scaling behaviour
of the number of conformations and their average spatial extent as a
function of the number of monomers and the associated critical
exponents $\gamma$ and $\nu$ are examined with two complementary
approaches: numerical chain-growth computer simulations using the
PERM algorithm and complete enumerations of all possible polymer conformations
employing a recently developed very efficient exact counting method.

\pacs 64.60.al, 07.05.Tp, 64.60.ah 
\end{abstract}

\section{Introduction}

Polymers in chemical and biological physics are often characterized
by a linear chemical architecture and behave as (semi-)flexible chains.
Typical examples of flexible polymers are synthetic polymers with a
carbon backbone, such as polyethylene, where the carbon bonds along
the backbone can easily rotate against each
other \cite{Flory1953,Gennes,desCloizeaux90}.
Recently, much interest has been paid to semiflexible polymers, mostly because important biopolymers such as DNA, actin filaments in eukaryotic
cells and some other proteins belong to this class \cite{Bustamante94,Bustamante94a}.
Typically, such biopolymers are supramolecular assemblies with a relatively
large monomer diameter representing many atoms. Some synthetic polymers
also exhibit bending stiffness over short distances along the chain. The
competition between thermal energy and the bending energy of the polymer
sets a characteristic length scale, the persistence length $l_\textrm{p}$, describing
the crossover between two main regimes: for chain lengths much larger than $l_\textrm{p}$,
any polymer behaves as a flexible chain,
whereas for chain lengths much
smaller than the persistence length, the polymer attains the limit of a
rigid rod.

The conformational properties, that is size and shape of polymer coils,
play an important role in models explaining the viscous flow and other
hydrodynamic properties of polymer fluids. It is well established
\cite{Flory1953,Gennes,desCloizeaux90} that under good solvent conditions
the conformation of a flexible chain can be characterized by a number
of universal properties. This enables one to classify a broad variety
of polymers, independently of their precise chemical structure, into a
single universality class in space dimension $d$. Physically relevant
cases are both polymers in three spatial ``bulk'' dimensions, $d=3$, and
polymers which are completely adsorbed onto a planar substrate and hence
confined to two dimensions. A simple example is the averaged radius of
gyration $R_\textrm{G}$, which defines the effective size of a typical $N$-monomer
chain and obeys the scaling law
 \begin{equation}
	  \langle R_\textrm{G}^2 \rangle \sim  N^{2\nu}, \label {RR}
\end{equation}
where the universal exponent $\nu$ takes on the value
$\nu = 3/4$ in $d=2$ dimensions \cite{Nienhuis82} and
$\nu=0.587\,597 \pm 0.000\,007$ in $d=3$ dimensions \cite{Clisby2010} for
different flexible polymers
in a good solvent.
Another important conformational characteristic is the universal averaged
asphericity  $A$  \cite{Aronovitz86}, distinguishing completely stretched,
rod-like conformations ($A=1$) from spherical ones ($A=0$).
For flexible polymers in good solvents, independent of their chemical
structure, $A=0.431\pm0.002$ in $d=3$ \cite{Bishop88}, indicating a
significant anisotropy of polymer coil conformations. Due to universality,
the conformational properties of flexible macromolecules mentioned above
are perfectly captured within the lattice model of self-avoiding walks
(SAWs) \cite{Vanderzande98}.

In general, the overall size and shape of a complex polymer macromolecule are
controlled by the monomer-monomer interactions, usually including
van der Waals attraction and, if the constituents are charged, also long-range
Coulomb interaction. In continuum models, the hard-core repulsion and van der Waals
attraction can be modelled by Lennard-Jones potentials. Employing a lattice
discretization of space, one considers the so-called interacting self-avoiding
walks (ISAWs) (sometimes also denoted as ``self-attracting self-avoiding
walks (SASAWs)''), where apart from the self-avoidance constraint, representing
the hard-core potential, the van der Waals contribution is modelled by a
nearest-neighbour interaction,
\begin{equation}
	H = -\epsilon N_i .
\end{equation}
Here the parameter $\epsilon$ sets the energy scale and $N_i$ counts the
number of contacts of the polymer with itself. This may be depicted as a
step-like potential which qualitatively corresponds
to a Lennard-Jones interaction with large-distance cutoff.
Such polymer models have been applied to a large variety of problems including
protein folding (where $\epsilon$ depends on the type of interacting residues)
and surface
adsorption \cite{Lau89,Janke03a,Janke03b,Schiemann,mb+wj-adsorb1,mb+wj-adsorb2,mb+wj-adsorb3,Bachmann08,zaragoza08}.
They provide a coarse-grained approximation to flexible
$\Theta$-polymers and are a suitable and very efficient way for studying
generic properties of the collapse and freezing transitions for relatively long
polymer chains.

In real physical processes, this ideal picture is further complicated by the
fact that structural obstacles (impurities) in the environment may
alter the behaviour of the system. The density
fluctuations of obstacles may lead to a large spatial inhomogeneity and
create pore spaces, which are frequently of fractal structure \cite{Dullen79}.
%
%
The understanding of the behaviour of macromolecules in porous
environments, that is in the presence of structural disorder,
 e.g., in colloidal solutions or microporous membranes \cite{Cannel80}, is of
 great importance in polymer physics.
In particular, a related problem is relevant when studying
the protein folding dynamics in a cellular environment. Biological cells can
be described as a highly disordered (``crowded'') environment due to the
presence of a large amount of soluble and insoluble biochemical species,
which can be estimated to occupy up to $40\%$ of the total aquabased
volume \cite{Minton01a,Minton01b}. It is known that structural obstacles strongly affect protein folding
and aggregation \cite{Horwich}. Here, we will model
the disordered environment generically by forcing the SAWs to only walk
on the sites of percolation clusters at the threshold
value \cite{Barat95,Kumar-Li} where they exhibit a self-similar, fractal structure \cite{Stauffer}.

As established by de Gennes \cite{Gennes}, the scaling properties of
infinitely long flexible polymers can be evaluated by analyzing the critical
behaviour of the $n$-component vector spin model in the formal $n \to 0$
limit (so-called polymer limit). In particular, the polymer size exponent
$\nu$ as defined in the scaling law (\ref{RR}) is related by universality to
the correlation length critical index
of the $n=0$-component model, whereas the universal shape parameters of polymer
chains can be computed by field theoretic methods in terms of the critical
amplitude ratios of this
model \cite{Aronovitz86}. Correspondingly, the problem of polymers in disordered
environments can be mapped to analyses of the critical behaviour of the randomly
diluted $n$-vector model in the polymer limit \cite{Kim83,Blavatska01}.

Another class of problems concerns the behaviour of polymers in the presence
of a planar substrate, which can be either energetically neutral (presenting
a purely geometrical constraint), or attractive for the polymer
\cite{Eisenriegler82,Eisenriegler93,mb+wj-adsorb1,mb+wj-adsorb2,mb+wj-adsorb3,Bachmann-a,Bachmann-b,Bachmann-c,Bachmann-d,Bachmann-e}. In the latter case,
the polymer will be adsorbed to the surface at low temperatures and desorbed at
high temperatures.
The adsorption of polymers on surfaces is a long-standing problem which plays
an important role in a number of technological applications (lubricants,
stabilization of colloids) and biology
(adsorption of biopolymers onto membranes) \cite{Xie02}.
For example, much interest has been paid recently to experimental studies of the adsorption of peptides
onto semiconductor substrates \cite{Whaley00, Goede10},
being of particular importance for future sensory devices and pattern
recognition at the nanometer scale.

A subject of particular interest is the adsorption of polymers onto disordered (fractal) surfaces, since naturally occurring substrates
are usually rough and energetically (or structurally) inhomogeneous \cite{avnir84a,avnir84b}, and the fractal surface geometry is known to have a crucial effect on adsorption
phenomena \cite{Kawaguchi91,Huber98}.
Already simple physical arguments suggest that by increasing
the surface irregularity, the number of polymer-surface contacts is strongly
altered relative to the idealized planar surface without being balanced by a
loss in configurational entropy. This is a consequence of changes in the
probability of polymer-surface contacts with increasing roughness. A
fractal surface is in general characterized by its fractal dimension $d_\textrm{s}$
($d_\textrm{s}=2$ corresponds to a flat surface). In particular, it was
shown that adsorption is enhanced (diminished),
when the fractal dimension of the surface is larger (smaller) than~2 \cite{Bouchaud89}.

The rest of this mini-review is organized as follows. In the next section,
the employed methods, PERM chain-growth computer simulations and exact enumerations,
are described and their merits and drawbacks compared.
Section~3 contains an overview of our results for the scaling behaviour of
SAWs on ``bulk'' percolation clusters and a shorter paragraph on the
adsorption properties to a fractal surface. Finally, in section~4 we finish with
some concluding remarks and give an outlook to ongoing and future work.

\section{Methods}

\begin{figure}[!t]
\begin{center}
\includegraphics[width=0.5\textwidth]{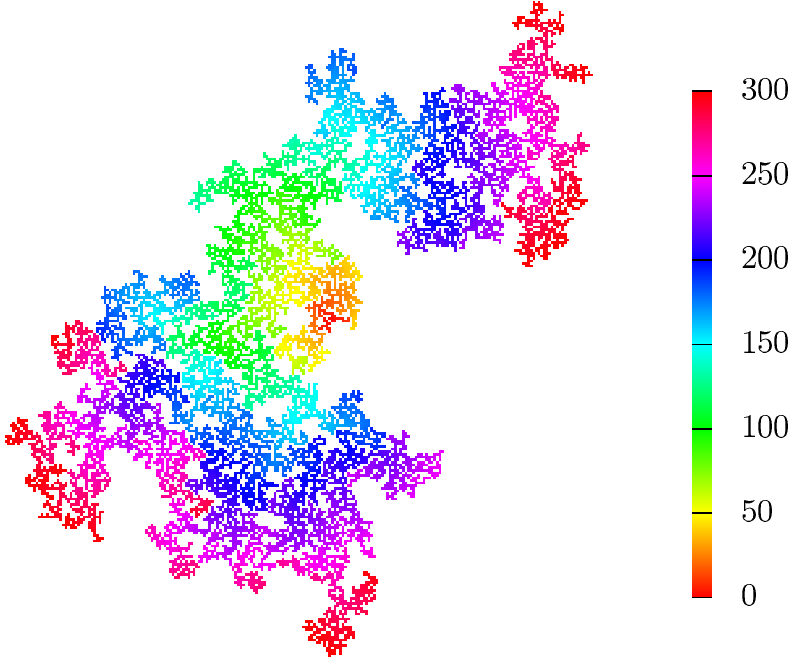}
\caption{(Color online) Example of a critical percolation cluster in two dimensions.
	Colors show the chemical distance to the origin, which is in the red center.
	Sites more than $300$ steps away are not shown.}
\label{cluster}
\end{center}
\end{figure}

Whereas for continuum polymer models Markov chain Monte Carlo (MC) and
Molecular Dynamics (MD) simulations are the methods of
choice \cite{wj-lnp,wj-lnp2,wj-ising-lect}, for lattice
polymer models alternative numerical techniques may be employed that are better
tailored to the problem at hand. The methods in a first
class, chain-growth algorithms, are based on stochastic sampling,
similarly to standard MC methods. In certain situations, some variants are also suitable for continuum
studies \cite{GarelOrland,schoebl,schoebl2}. The second type, complete or exact
enumerations, on the other hand, explicitly exploit that the configuration
space of lattice polymers is discrete. Complete enumerations yield exact averages of the polymer statistics for a
given lattice structure. The numerical
complexity of the method, however, grows in general exponentially with the
number of monomers so that only relatively short chains (with of the order
of $30\div50$ monomers) can be investigated by this means. In the next two
subsections, first the specific chain-growth algorithm ``PERM'' will be
briefly explained. Subsequently, for the special case of critical percolation
clusters at the percolation threshold as depicted in figure~\ref{cluster}, it
is shown how the exponential complexity of complete enumerations can be reduced
to a polynomial one when their self-similar fractal structure is exploited in
a suitable way.

\subsection{PERM chain-growth algorithm}

The algorithm underlying the pruned-enriched Rosenbluth method
(PERM) \cite{Grassberger97} is based on ideas from the very
first days of Monte Carlo simulations, the Rosenbluth-Rosenbluth (RR)
method \cite{Rosenbluth55} combined with enrichment strategies \cite{Wall59}.
To grow a polymer chain, one places the $n$th monomer at a randomly chosen
nearest-neighbour site of the last placed $(n-1)$th  monomer and continue
iteratively until the total length of the chain $n = N$ is reached.

In order to obtain correct statistics, any attempt
to place a new monomer at an already occupied site
results in discarding the entire chain. This leads to
the problem of an exponential ``attrition'', which means that
the number of discarded
chains grows exponentially with the chain length $N$.  Clearly, this renders the
method useless for long chains. In the RR method, occupied neighbours are avoided
without discarding the chain. This would introduce a bias in the proper statistics
which, however, can be corrected by giving a weight $W_n\sim \prod_{l{=}2}^n m_l$
to each sample conformation in the $n$th step, where $m_l$ is the number of free
lattice sites to place the $l$th monomer. When an ensemble of chains of total
length $N$ is constructed, any new chain has to start from the same starting point,
until the desired number of chain conformations is obtained. The configurational
averaging for, e.g., the end-to-end distance $r\equiv \sqrt{R^2(N)}$ is then given by
\begin{eqnarray}
	&&\langle {r} \rangle {=} \frac{\sum_{\textrm{conf}} W_N^{\textrm{conf}} r^{\textrm{conf}}}{\sum_{\textrm{conf}} W_N^{\textrm{conf}}}{=}\sum_{r}rP(r,N) \label{R} \,,
\end{eqnarray}
where $W_N^{\textrm{conf}}$ is the Rosenbluth weight of an $N$-monomer chain in a given
conformation and $P(r,N)$ is the distribution function for the end-to-end distance.

While the chain grows by adding monomers, its Rosenbluth weight will fluctuate.
PERM suppresses
these fluctuations by a population control, namely by ``pruning'' conformations with
too small weights and by ``enriching'' the sample with copies of high-weight
conformations \cite{Grassberger97}.
These copies are made while the chain is growing and continue to grow independently
of each other. Pruning and enrichment are performed by choosing thresholds $W_n^{<}$
and $W_n^{>}$ depending on the estimate of the partition sum
$Z_n = \sum_\textrm{conf} W_n^\textrm{conf}$ of the $n$-monomer chain.
These thresholds are continuously updated as the simulation progresses. The zeroth
iteration is a pure chain-growth algorithm without reweighting. After the first
chain of full length has been obtained, we switch to $W_n^{<}$, $W_n^{>}$. If
the current weight $W_n$ of an $n$-monomer chain is less than $W_n^{<}$, a
random number $r{=}{0,1}$ is chosen; if $r{=}0$, the chain is discarded, otherwise
it is kept and its weight is doubled. Thus, low-weight chains are pruned with
probability $1/2$. If $W_n$ exceeds $W_n^{>}$, the conformation is doubled and
the weight of each copy is taken as half the original weight. To update the
threshold values, one may apply similar empirical rules as in \cite{Hsu03,Janke03a,Janke03b}:
$W_n^{>}{=}C(Z_n/Z_1)(c_n/c_1)^2$ and $W_n^{<}{=}0.2W_n^{>}$, where $c_n$ denotes
the number of the created chains having length $n$, and the parameter $C \leqslant 1$ controls
the pruning-enrichment statistics. After a certain number of chains of total
length $N$ is produced, the iteration is finished and a new so-called tour starts.

\subsection{Exact enumeration}

The straightforward exact enumeration method for SAWs is to use brute force
and generate all possible conformations; see~\cite{,Lam1990,Ordemann2000,Singh2009}.
However, the computational work load becomes rapidly prohibitively high since
the number of walks $Z$ increases exponentially with the number of steps $N$,
\begin{equation}
	Z \simeq \mu^N N^{\gamma-1} = N^{\gamma-1} \re^{N\ln\mu} \,.
	\label{eq:gamma}
\end{equation}
Here, $\gamma$ is another universal exponent depending only on the dimension
and $\mu$ is the non-universal so-called connectivity constant which besides the dimension
does also depend on the details of the considered lattice.

\begin{figure}[!t]
\centering
\includegraphics[scale=2.1]{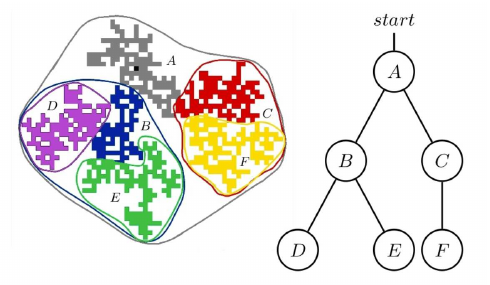}
\caption{(Color online) Decomposition of a critical percolation cluster into nested blobs and the
corresponding tree hierarchy. The starting position of the SAW is marked black
(in blob $A$).}
\label{decomp}
\end{figure}

Fortunately though, we
recently discovered that the fractal structure of percolation clusters offers a
way to circumvent this problem~\cite{Fricke2012,Fricke2012b}. The actual
implementation of our method is rather complicated and will not be explained
here in detail, but the
basic ideas are fairly simple. The key lies in the observation that the critical
clusters are very weakly connected, so that they could be divided by cutting only
a small number of bonds. Thanks to the self-similarity, this applies on all
length scales. We can, therefore, partition the cluster into a hierarchy of
nested ``blobs'' with very few interconnections in order
to {\em factorize\/} the enumeration (see figure~\ref{decomp}). This factorization
property is the main clue of the method.

\begin{figure}[!t]
\centering
\includegraphics[scale=0.7]{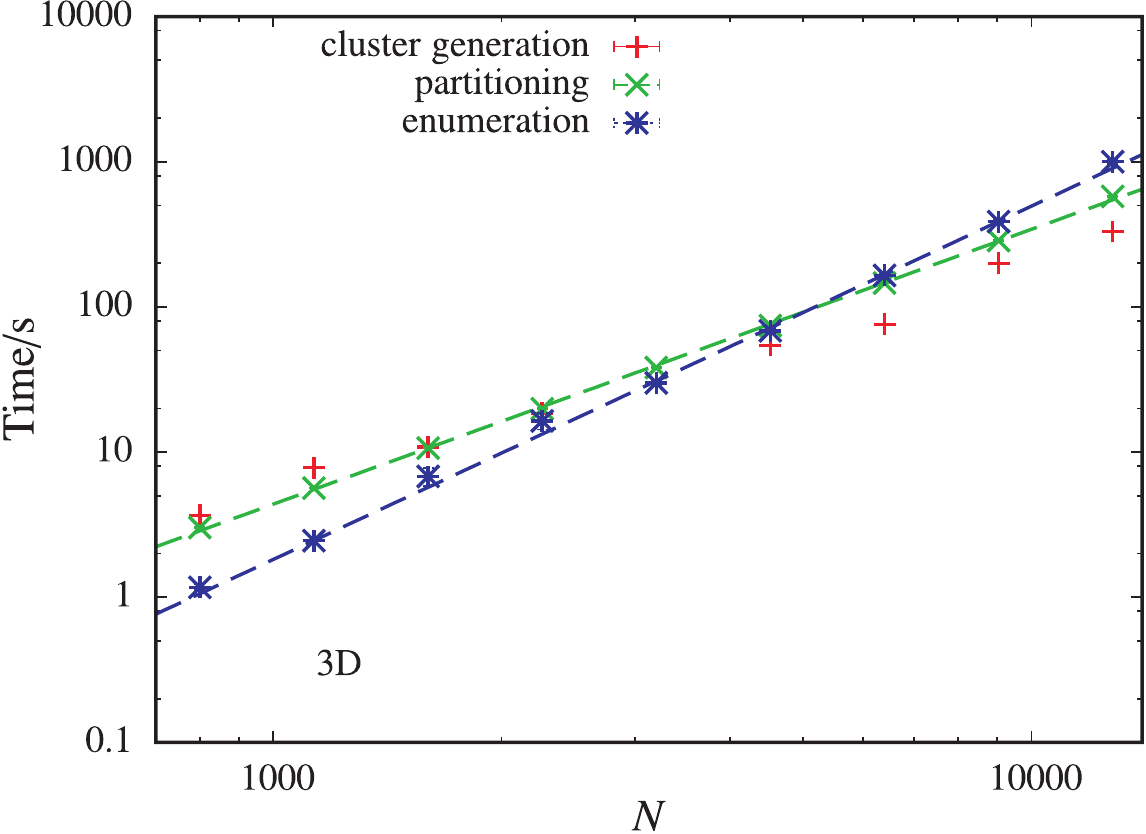}
\caption{(Color online) Computation time (in seconds on a typical 3 GHz PC) needed for the
complete enumeration of all conformations of an $N$-step self-avoiding walk on a
critical percolation cluster in three dimensions using our recently developed
recursive exact enumeration algorithm. The straight lines in the log-log plot
clearly indicate a power-law dependence on $N$ instead of the exponential
increase for standard enumeration techniques.
}
\label{times3D}
\end{figure}

We start by enumerating all SAW segments of any length through the smallest
blobs and divide them into different classes, depending on how they connect
the different entries to the blob. For instance, if a blob has two
connections ($a$ and $b$) to its parent in the hierarchy, we distinguish
between the following classes: segments starting at $a$ and terminating
within the blob, segments starting at $b$ and terminating within the blob, and
segments connecting $a$ and $b$. For the terminating segment classes, we also
measure the mean distance of the chain end to the origin. We then ``renormalize'' the
smallest blobs, treating them essentially as point-like, while we repeat the
procedure for the next larger blobs that contain them. The new segments should
then be matched with the right segment classes from before to determine
their multiplicity and average end-point distances. This scheme is applied
repeatedly going to ever larger blobs and ultimately the whole percolation cluster.
Correctly implemented, this method achieves polynomial (rather that exponential)
increase of computation time with the number of steps, thus allowing for
walks of several thousand steps. In fact, we can exactly enumerate as many as
$10^{1000}$ chain conformations~--- which with standard techniques would take
very long indeed (well, with current computers much longer than the estimated
age of our Universe \dots). Our recursive algorithm can be used in any dimension
as is illustrated in figure~\ref{times3D} for the three-dimensional case,
but we have only recently generalized our computer implementation to more than
two dimensions \cite{niklas_tobe,60a}, so that results for physical quantities on percolation clusters in three
or more dimensions are still somewhat premature.

The efficiency of our recursive enumeration method depends on the self-similar,
fractal blob-like structure of the critical percolation clusters, allowing
a factorization of the counting problem. Away from the
percolation threshold, its performance deteriorates and PERM chain-growth
simulations may be superior. Also, for measuring more detailed shape characteristics,
           such as the full gyration tensor and related universal
shape invariants \cite{Blavatska10,Blavatska11a} as well as for a
generalization to $\Theta$-polymers \cite{Blavatska09b,Blavatska11a},
PERM is more flexible. So in this sense the two methods are complementary
to each other.

\section{Results}

Let us start with our results for the scaling properties of SAWs
on $d$-dimensional critical percolation
clusters \cite{Blavatska08a,Blavatska08b,Blavatska08c,viktoria_jpa,Blavatska10b}.
To be precise, in the PERM
simulations we rather considered the (geometric) backbone of the cluster
which is the subset of the cluster consisting of all bonds (or sites) through
which a ``current'' can flow, i.e., it is the structure left when all ``dangling
ends'' are eliminated from the
cluster. The usual
argument is that SAWs can
be trapped in such ``dangling ends'', so that infinitely long chains can
only exist on the backbone of the cluster.

In the PERM simulations, we adjusted the pruning-enrichment control
parameter such that on average 10 chains of total length
$N$ ($ = 90, 80, 70$ in $d = 2,3,4$)
are generated per iteration \cite{Janke03a,Janke03b}, and performed
$10^6$ iterations. Also, what is even more important for efficiency,
we made sure that in almost all iterations at least one such a chain
was created. By comparison with our exact enumeration results for randomly
selected clusters, we have recently checked \cite{Fricke2013} that once
PERM has managed to squeeze the chains through the bottlenecks of the
percolation cluster, the convergence to the exact result is very fast.
Initially, however, PERM needs a sufficient period of time to explore
the landscape and to adjust the thresholds of the population control.

In the given problem we have to perform two types of averaging: The first
average is performed over all SAW conformations created with PERM on a
single backbone according to (\ref{R}), and the second average is carried
out over different realizations of disorder,
\begin{equation}
\overline{\langle r \rangle}{=}\frac{1}{N_\textrm{c}}\sum_{i{=}1}^{N_\textrm{c}} \langle r\rangle_i \,,
\label{eq:dis_av}
\end{equation}
where $N_\textrm{c}$ is the number of different clusters. In our PERM simulations we took
$N_\textrm{c} = 1000$.

\begin{figure}[!t]
\begin{center}
\includegraphics[width=0.46\textwidth]{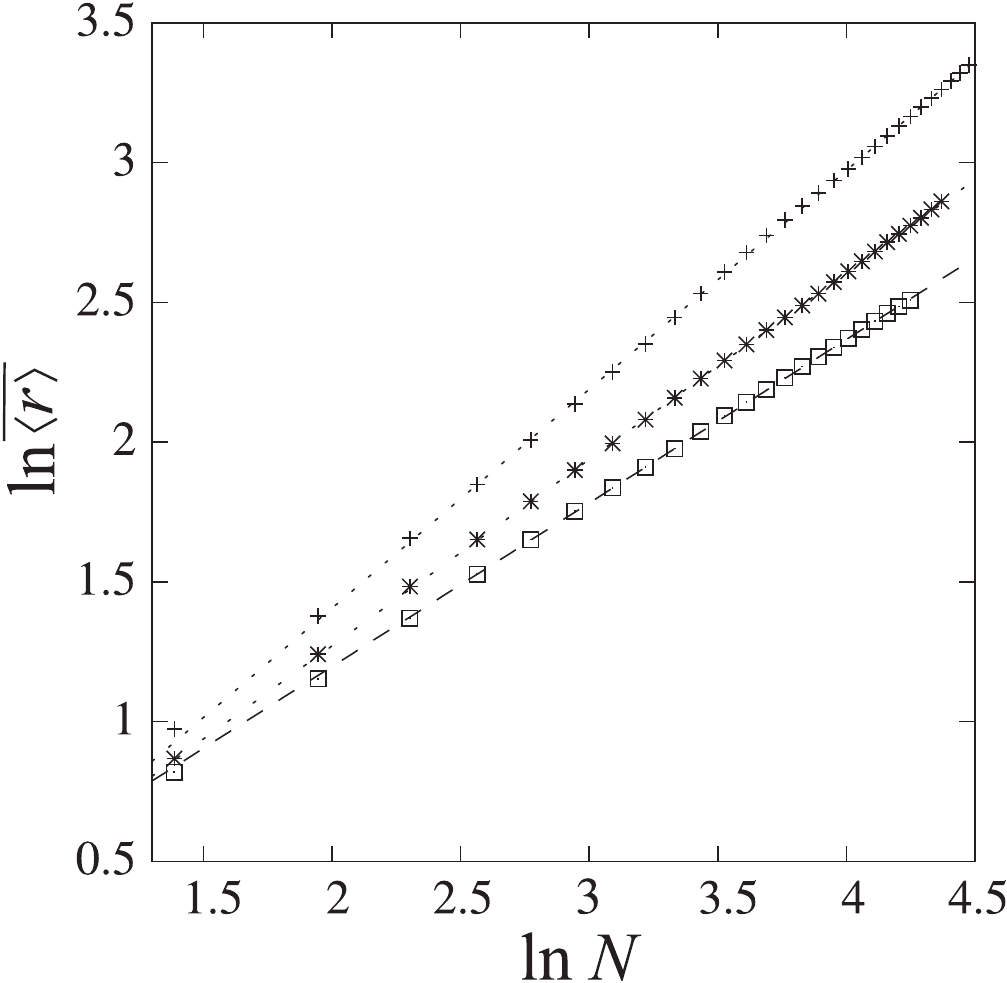}
\caption{Disorder averaged  end-to-end distance vs the number of steps
obtained in PERM chain-growth simulations
in double logarithmic scale for SAWs on the backbone of percolation clusters in
$d{=}2$ (pluses), $d{=}3$ (stars), $d{=}4$ (open squares).
Lines represent linear fitting, statistical error bars are of the size of symbols.}
\label{sawr}
\end{center}
\end{figure}

In this way we arrived at the data for the end-to-end distance shown
in figure~\ref{sawr} \cite{Blavatska08a}. To estimate the critical
exponent $\nu_{p_\textrm{c}}$,
we employed in a log-log representation linear least-square fits with
varying lower cutoff for the number of steps $N_\textrm{min}$. The $\chi^2$
value (sum of squares of normalized deviation from the regression line)
serves as a test of the goodness of fit and thus determines fit ranges
where the leading scaling ansatz (\ref{RR}) is acceptable. The resulting
estimates of the exponent $\nu_{p_\textrm{c}}$ \cite{Blavatska08a}
are compiled in the second last line
of table~\ref{allnu},
where previously obtained
results \cite{Rintoul94,Blavatska04,Janssen07,Woo91,Grassberger93,Lee96}
are listed also for comparison.

\begin{table}[!b]
\caption{ \label{allnu} Estimates of the exponent $\nu_{p_\textrm{c}}$ for SAWs on
critical percolation clusters.
EE: exact enumerations,
RG: field-theoretic renormalization group,
MC: Monte Carlo simulations.
For SAWs on the regular lattice one has:
$\nu_{{\rm SAW}}(d{{=}}2){=}3/4$~\cite{Nienhuis82},
$\nu_{{\rm SAW}}(d{{=}}3){=}0.587\,597 \pm 0.000\,007$~\cite{Clisby2010},
$\nu_{{\rm SAW}}(d\geqslant 4){=}1/2$.}
\vspace*{3mm}
\begin{center}
\begin{tabular}{|r| c| c|  c|  }
\hline\hline $\nu_{p_\textrm{c}} \setminus d$  & 2 & 3 & 4 \\
\hline\hline
EE \cite{Rintoul94}&$0.770 \pm 0.005$& $0.660 \pm 0.005$&\\
\cite{Ordemann2000}&$0.778 \pm 0.015$& $0.66 \pm 0.01$& \\
\cite{Ordemann2000}&$0.787 \pm 0.010$& $0.662 \pm 0.006$&\\
\hline
RG \cite{Blavatska04} & 0.785 & 0.678& 0.595 \\
\cite{Janssen07} & 0.796 & 0.669& 0.587 \\
 \hline MC
 \cite{Woo91}
  & $0.77  \pm 0.01$  & &   \\
  \cite{Grassberger93} & $0.783 \pm 0.003$ & &
  \\
  \cite{Lee96} & & 0.62--0.63 &0.56--0.57
 \\
 \hline
 PERM \cite{Blavatska08a} & $ 0.782\pm 0.003$ & $0.667\pm 0.003$ & $0.586\pm 0.003$ \\
 EE \cite{Fricke2012} & $ 0.7754\pm 0.0015$ & &
 \\
 \hline\hline
 \end{tabular}
 \end{center}
\end{table}

With our new exact enumeration scheme, we have so far only analyzed the
two-dimensional case \cite{Fricke2012,Fricke2012b}. Our result for $\nu_{p_\textrm{c}}$
in the last line of table~\ref{allnu} is
based on walks with up to $N=1\,000$ steps averaged over $200\,000$ percolation
clusters at $p_\textrm{c}$ (working here with the full cluster, not the backbone).
In the least-square fit to the scaling law (\ref{RR}), the first 500 steps were
excluded to reduce finite-size effects. In the meantime we have extended
our exact enumeration program to up to seven dimensions, and by several technical
refinements of the actual computer implementation we can now handle walks with
over $10\,000$ steps exactly. The data analysis, however, is still in progress and
will be reported elsewhere \cite{niklas_tobe,60a}.

Estimates of the exponent $\gamma$ in (\ref{eq:gamma}) are much more difficult to
obtain since $\gamma$ appears in the subleading term, which in the present context
is a severe problem because
$Z$ exhibits strong fluctuations from cluster to cluster. In fact,
the distribution of $Z$ has log-normal character due to the fact that $Z$
is effectively a product of random variables. Still, performing least-square fits to
\begin{equation}
\frac{\ln \overline{Z}}{N} = \frac{\ln A}{N} + \ln \mu +(\gamma-1)\frac{\ln N}{N} \,,
\label{eq:gamma_fit}
\end{equation}
we obtain for instance in two dimensions from PERM simulations \cite{viktoria_jpa}
the estimates $\mu = 1.566 \pm 0.005$, $\gamma = 1.350 \pm 0.008$ and from the exact
enumerations \cite{Fricke2012} (in the interval $N = 10-50$) $\mu = 1.567 \pm 0.003$,
$\gamma = 1.35 \pm 0.03$, which are in surprisingly close agreement. Note that
$\mu$ is basically given by the product of the connectivity constant for the
regular square lattice and the percolation threshold ($2.6385 \times 0.592\,746
= 1.5639$) and that $\gamma$ is hardly distinguishable from its value for the
{\em regular\/} two-dimensional lattice ($\gamma = 43/32 = 1.3437\dots$).
With PERM we have estimated $\mu$, $\gamma$ also in three and four dimensions
\cite{viktoria_jpa}, whereas the exact enumeration data still have to be
analyzed \cite{niklas_tobe,60a}.

As an interesting side result, we obtained with PERM also numerical estimates
for the averaged asphericity, prolateness, and size ratio of the chains
in two and three dimensions
\cite{Blavatska10}. All these shape characteristics
can be expressed in terms of combinations of the components of the gyration
tensor. They increase gradually with increasing polymer chain length~--- the
fractal structure of the percolation cluster drives the longer polymer chain
conformations to become more and more prolate. Our results quantitatively
indicate that the shape parameters of typical polymer conformations change
significantly relative to the obstacle-free case. For example, we obtained for
the (universal) asphericity of SAWs in three dimensions
$\langle A_3 \rangle = 0.435 \pm 0.002$
on a regular lattice in good agreement with \cite{Bishop88} and
$\overline{\langle A_3^{p_\textrm{c}} \rangle} = 0.743 \pm 0.005$
on critical percolation clusters: The shape tends to be more anisotropic
and elongated due to the fractal structure of the disordered environment.

\begin{figure}[!t]
\begin{center}
\includegraphics[width=0.48\textwidth]{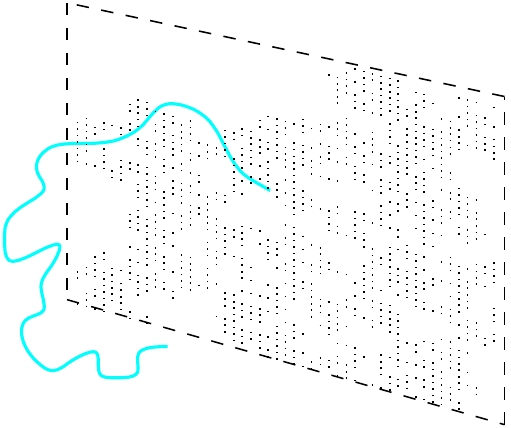}
\caption{
\looseness=-1
(Color online) Sketch of a polymer chain grafted with one end to a randomly
           selected site of an attractive two-dimensional percolation cluster modelling a fractal surface with
fractal dimension $d_\textrm{s}^{p_\textrm{c}}=91/48 \approx 1.896 < 2$.\vspace*{-2mm}
\label{fig:mem}}
\end{center}
\end{figure}

Let us finally briefly comment on the conformational statistics of polymers
interacting with disordered substrates.
Particularly interesting is the case
of an attractive substrate, where below a critical temperature $T_A$, a
second-order phase transition into an adsorbed state takes place. As order
parameter, one considers the fraction $N_\textrm{s}/N$, where $N_\textrm{s}$ is the average number
of monomers adsorbed to the surface and $N$ is the length of the
polymer chain. In the limit of long chains, it obeys the scaling law
$\langle N_\textrm{s} \rangle/N \sim N^{\phi_\textrm{s}-1}$, where $\phi_\textrm{s}$ is the so-called
surface crossover exponent \cite{Eisenriegler82}.
In the language of lattice models,
disordered surfaces can be modelled as a two-dimensional
regular lattice with different types of randomly distributed defects as
sketched in figure~\ref{fig:mem}. Fractal properties emerge at the (two-dimensional)
percolation threshold where a spanning percolation cluster of attractive sites
with fractal dimension\footnote{Please note that in references \cite{Blavatska08a,Blavatska08b,viktoria_jpa,viktoria_surface,viktoria_surface_athens}
inadvertently a wrong value of 91/49 has been quoted. In table~2 of reference \cite{viktoria_jpa},
           this led to an incorrect value of 7/9 ($\approx 0.7777\dots$) for
           a Flory-type approximation, which properly should read
           144/187 \linebreak ($\approx 0.7700\dots$).} $d_\textrm{s}^{p_\textrm{c}}=91/48 \approx 1.896 < 2$ appears \cite{Havlin87}.

In reference \cite{viktoria_surface} we have studied this problem with the help of
PERM. We examined the behaviour of the components of the radius of gyration
$\overline{\langle R^2_{\textrm{g}{||}}\rangle}$, $\overline{\langle R^2_{\textrm{g}{\perp}}\rangle}$
in directions parallel and perpendicular to the surface and found that
the critical exponent governing the scaling of the size of the polymer chain
adsorbed on the fractal substrate
is larger than that for a homogeneously attractive surface. A value
$\nu_2^{p_\textrm{c}}=0.772\pm0.006$ is obtained \cite{viktoria_surface,viktoria_surface_athens}, to be compared with
the compatible result $\nu_2^{p_\textrm{c}} = 0.782 \pm 0.003$ for the average size
of a polymer strictly confined onto a two-dimensional percolating
cluster \cite{Blavatska08a,viktoria_jpa} and $\nu_2 = 0.742 \pm 0.006 \approx 3/4$
for a plain surface.
Examining the peak structure of the heat capacity, we obtained an estimate for
the surface crossover exponent of $\phi_\textrm{s}^{p_\textrm{c}} = 0.425 \pm 0.009$,
compared to $\phi_\textrm{s} = 0.509 \pm 0.009$ for the plain
surface \cite{viktoria_surface,viktoria_surface_athens}. As
expected \cite{Bouchaud89}, the adsorption is diminished when the fractal
dimension of the surface is smaller than that of the plain Euclidean surface
due to the smaller number of contacts of monomers with attractive sites.

\clearpage

\section{Concluding remarks and outlook}

The scaling behaviour of polymers in disordered environments is a longstanding
problem that has been investigated in many publications over the last 25 years.
At a closer look, however, there still remain many open questions and
controversial issues. In most numerical studies so far, for technical reasons
the length $N$ of the polymer chains has been quite short, so that it is not
guaranteed at all that the true asymptotic behaviour has been observed. This
issue is for instance important when discussing whether SAWs on the backbone
of a critical percolation cluster exhibit the same critical exponents as on
the full cluster. Our recently further improved
exact enumeration scheme \cite{niklas_tobe} allows us now for the first time
to study the chains in up to $d=7$ dimensions with over $10\,000$ monomers exactly
and hence offers a completely new perspective. Our preliminary data analysis
indeed suggests that new insights into an old problem may be expected in the
near future.

It is also possible to generalize our enumeration method to $\Theta$-polymers
in disordered environments which enables us to study with unprecedented accuracy the collapse and freezing
transitions of IASWs on critical percolation clusters.
A stretching force \cite{Kumar-Li,kumar-stretching,kumar-stretching2,Singh2009,Blavatska09b,Blavatska11a}
can also be taken into account. Finally, a careful comparison with related random walk types, e.g., kinetic growth walks \cite{niklas-johannes},
may shed new light on the rather involved statistical properties of polymers in
disordered environments.

\section*{Acknowledgements}

This work has been supported by an Institute Partnership Grant ``Leipzig--Lviv''
of the Alexander von Humboldt Foundation. Financial support by
the S\"achsische DFG Forschergruppe FOR877 under Grant No.~JA 483/29-1
and the
DFG Sonderforschungsbereich SFB/TRR 102 (project B04)
is gratefully acknowledged.


\begin{thebibliography}{99}

\bibitem{Flory1953}
Flory P., Principles of Polymer Chemistry, Cornell University Press,
Ithaca, 1953.

\bibitem{Gennes}
de~Gennes P.-G., Scaling Concepts in Polymer Physics, Cornell University Press,
Ithaca and London, 1979.

\bibitem{desCloizeaux90}
Des Cloizeaux J., Jannink G., Polymers in Solution, Clarendon Press, Oxford, 1990.

\bibitem{Bustamante94}
Bustamante C., Marko J.F., Siggia E.D., Smith S., Science, 1994, \textbf{265}, 1599; \doi{10.1126/science.8079175}.

\bibitem{Bustamante94a}
Ober C.K., Science, 2000, \textbf{288}, 448; \doi{10.1126/science.288.5465.448}.


\bibitem{Nienhuis82}
Nienhuis B., Phys. Rev. Lett., 1982, \textbf{49}, 1062; \doi{10.1103/PhysRevLett.49.1062}.

\bibitem{Clisby2010}
Clisby N.,
Phys. Rev. Lett., 2010, \textbf{104}, 055702; \doi{10.1103/PhysRevLett.104.055702}.

\bibitem{Aronovitz86}
Aronovitz J.A., Nelson D.R., J. Phys. France, 1986, \textbf{47}, 1445; \doi{10.1051/jphys:019860047090144500}.

\bibitem{Bishop88}
Bishop M., Saltiel C.J., J. Chem. Phys., 1988, \textbf{88}, 6594; \doi{10.1063/1.454446}.

\bibitem{Vanderzande98}
Vanderzande C., Lattice Models of Polymers, Cambridge University Press,
Cambridge, 1998.

\bibitem{Lau89}
Lau K.F., Dill K.A., Macromolecules, 1989, \textbf{22}, 3986; \doi{10.1021/ma00200a030}.

\bibitem{Janke03a}
Bachmann M., Janke W.,
Phys. Rev. Lett., 2003, \textbf{91}, 208105; \doi{10.1103/PhysRevLett.91.208105}.

\bibitem{Janke03b}
Bachmann M., Janke W.,
J. Chem. Phys., 2004, \textbf{120}, 6779; \doi{10.1063/1.1651055}.

\bibitem{Schiemann}
Schiemann R., Bachmann M., Janke W.,
J. Chem. Phys., 2005, \textbf{122}, 114705; \doi{10.1063/1.1814941}.

\bibitem{mb+wj-adsorb1}
Bachmann M., Janke W.,
%
Phys. Rev. Lett., 2005, \textbf{95}, 058102; \doi{10.1103/PhysRevLett.95.058102}.

\bibitem{mb+wj-adsorb2}
Bachmann M., Janke W.,
%
Phys. Rev. E, 2006, \textbf{73}, 020901(R); \doi{10.1103/PhysRevE.73.020901}.

\bibitem{mb+wj-adsorb3}
Bachmann M., Janke W.,
%
Phys. Rev. E, 2006, \textbf{73}, 041802; \doi{10.1103/PhysRevE.73.041802}.

\bibitem{Bachmann08}
Bachmann M., Janke W., In: Rugged Free Energy Landscapes,
Janke W. (Ed.), Lecture Notes
in Physics Vol.~{736}, Springer, Berlin, 2008, 203--246; \doi{10.1007/978-3-540-74029-2_8}.

\bibitem{zaragoza08}
Bachmann M., Janke W.,
AIP Conf. Proc., 2008, \textbf{1071}, 1; \doi{10.1063/1.3033357}.


\bibitem{Dullen79}
Dullen A.L., Porous Media: Fluid Transport and Pore Structure, Academic,
New York, 1979.

\bibitem{Cannel80}
Cannell D.S., Rondelez F., Macromolecules, 1980, \textbf{13}, 1599; \doi{10.1021/ma60078a046}.

\bibitem{Minton01a}
Minton A.P., J. Biol. Chem., 2001, \textbf{276}, 10577; \doi{10.1074/jbc.R100005200}.

\bibitem{Minton01b}
Ellis R.J., Minton A.P., Nature, 2003, \textbf{425}, 27; \doi{10.1038/425027a}.

\bibitem{Horwich}
Horwich A., Nature, 2004, \textbf{431}, 520; \doi{10.1038/431520a}.

\bibitem{Barat95}
Barat K., Chakrabarti B.K., Phys. Rep., 1995, \textbf{258}, 378; \doi{10.1016/0370-1573(95)00009-6}.

\bibitem{Kumar-Li}
Kumar S., Li M.S., Phys. Rep., 2010, \textbf{486}, 1; \doi{10.1016/j.physrep.2009.11.001}.

\bibitem{Stauffer}
Stauffer D., Aharony A., Introduction to Percolation Theory, Taylor and Francis,
London, 1992.

\bibitem{Kim83}
Kim Y., J. Phys. C: Solid State Phys., 1983, \textbf{16}, 1345; \doi{10.1088/0022-3719/16/8/005}.

\bibitem{Blavatska01}
Blavats'ka V., von Ferber C., Holovatch Yu.,
Phys. Rev. E, 2001, \textbf{64}, 041102; \doi{10.1103/PhysRevE.64.041102}.

\bibitem{Eisenriegler82}
Eisenriegler E., Kremer K., Binder K.,
J. Chem. Phys., 1982, \textbf{77}, 6296; \doi{10.1063/1.443835}.

\bibitem{Eisenriegler93}
Eisenriegler E., Polymers Near Surfaces: Conformation Properties and Relation
to Critical Phenomena, World Scientific, Singapore, 1993.

\bibitem{Bachmann-a}
M\"oddel M., Bachmann M., Janke W., J. Phys. Chem. B, 2009, \textbf{113}, 3314; \doi{10.1021/jp808124v}.

\bibitem{Bachmann-b}
M\"oddel M., Janke W., Bachmann M., Phys. Chem. Chem. Phys., 2010, \textbf{12}, 11548; \doi{10.1039/c002862b}.

\bibitem{Bachmann-c}
M\"oddel M., Janke W., Bachmann M., Comput. Phys. Commun., 2011, \textbf{182}, 1961; \doi{10.1016/j.cpc.2010.12.016}.

\bibitem{Bachmann-d}
M\"oddel M., Janke W., Bachmann M., Macromolecules, 2011, \textbf{44}, 9013; \doi{10.1021/ma201307c}.

\bibitem{Bachmann-e}
M\"oddel M., Janke W., Bachmann M., Phys. Rev. Lett., 2014, \textbf{112}, 148303; \doi{10.1103/PhysRevLett.112.148303}.


\bibitem{Xie02}
Xie A.F., Granick S., Nat. Mater., 2002, \textbf{1}, 129; \doi{10.1038/nmat738}.

\bibitem{Whaley00}
Whaley S.R., English D.S., Hu E.L., Barbara P.F., Belcher A.M., Nature, 2000,
\textbf{405}, 665; \doi{10.1038/35015043}.

\bibitem{Goede10}
Bachmann M., Goede K., Beck-Sickinger A., Grundmann M., Irb\"ack A., Janke W.,
Angew. Chem. Int. Ed., 2010, \textbf{49}, 9530; \doi{10.1002/anie.201000984}.

\bibitem{avnir84a}
Avnir D., Farin D., Pfeifer P., J. Chem. Phys., 1983, \textbf{79}, 3566; \doi{10.1063/1.446211}.

\bibitem{avnir84b}
Avnir D., Farin D., Pfeifer P., Nature, 1984, \textbf{308}, 261; \doi{10.1038/308261a0}.

\bibitem{Kawaguchi91}
Kawaguchi M., Arai T., Macromolecules, 1991, \textbf{24}, 889; \doi{10.1021/ma00004a013}.

\bibitem{Huber98}
Huber G., Vilgis T.A., Eur. Phys. J. B, 1998, \textbf{3}, 217; \doi{10.1007/s100510050306}.

\bibitem{Bouchaud89}
Bouchaud E., Vannimenus J., J. Phys. France, 1989, \textbf{50}, 2931; \doi{10.1051/jphys:0198900500190293100}.

\bibitem{wj-lnp}
Janke W., In: Ageing and the Glass Transition, Henkel M., Pleimling M., Sanctuary R. (Eds.), Lecture Notes in Physics Vol.~{716}, Springer, Berlin, 2007,  207--260; \doi{10.1007/3-540-69684-9_5}.

\bibitem{wj-lnp2}
Janke W., In: Computational Many-Particle Physics,  Fehske H.,  Schneider R.,  Weisse A. (Eds.), Lecture Notes in Physics Vol.~739, Springer, Berlin, 2008, 79--140; \doi{10.1007/978-3-540-74686-7_4}.

\bibitem{wj-ising-lect}
Janke W.,
In: Order, Disorder and Criticality: Advanced Problems of
Phase Transition Theory, Vol.~3, Holovatch Yu. (Ed.),
World Scientific, Singapore, 2012, 93--166.

\bibitem{GarelOrland}
Garel T., Orland H.,
J. Phys. A: Math.  Gen., 1990, \textbf{23}, L621; \doi{10.1088/0305-4470/23/12/007}.

\bibitem{schoebl}
Sch\"obl S., Zierenberg J., Janke W.,
Phys. Rev. E, 2011, \textbf{84}, 051805; \doi{10.1103/PhysRevE.84.051805}.

\bibitem{schoebl2}
Sch\"obl S., Zierenberg J., Janke W.,
J. Phys. A: Math. Theor., 2012, \textbf{45}, 475002; \doi{10.1088/1751-8113/45/47/475002}.

\bibitem{Grassberger97}
Grassberger P., Phys. Rev. E, 1997, \textbf{56}, 3682; \doi{10.1103/PhysRevE.56.3682}.

\bibitem{Rosenbluth55}
Rosenbluth M.N., Rosenbluth A.W., J. Chem. Phys., 1955, \textbf{23}, 356; \doi{10.1063/1.1741967}.

\bibitem{Wall59}
Wall F.T., Erpenbeck J.J., J. Chem. Phys., 1959, \textbf{30}, 634; \doi{10.1063/1.1730021}.

\bibitem{Hsu03}
Hsu H.P., Mehra V., Nadler W., Grassberger P., J. Chem. Phys., 2003, \textbf{118}, 444; \doi{10.1063/1.1522710}.

\bibitem{Lam1990}
Lam P.M.,
J. Phys. A: Math. Gen., 1990, \textbf{23}, L831; \doi{10.1088/0305-4470/23/16/010}.

\bibitem{Ordemann2000}
Ordemann A., Porto M., Roman H.E., Bunde A.,
Phys. Rev. E, 2000, \textbf{61}, 6858; \doi{10.1103/PhysRevE.61.6858}.

\bibitem{Singh2009}
Singh A.R., Giri D., Kumar S.,
Phys. Rev. E, 2009, \textbf{79}, 051801; \doi{10.1103/PhysRevE.79.051801}.

\bibitem{Fricke2012}
Fricke N., Janke W.,
Europhys. Lett., 2012, \textbf{99}, 56005; \doi{10.1209/0295-5075/99/56005}.

\bibitem{Fricke2012b}
Fricke N., Janke W.,
Physics Procedia, 2012, \textbf{34}, 39; \doi{10.1016/j.phpro.2012.05.006}.

\bibitem{niklas_tobe}
Fricke N., Janke W., Leipzig preprint (to be published).

\bibitem{60a}
Fricke N., Janke W., Preprint \arxiv{1409.3457}, 2014.

\bibitem{Blavatska10}
Blavatska V., Janke W.,
J. Chem. Phys., 2010, \textbf{133}, 184903; \doi{10.1063/1.3501368}.

\bibitem{Blavatska11a}
Blavatska V., Janke W.,
Comput. Phys. Commun., 2011, \textbf{182}, 1966; \doi{10.1016/j.cpc.2010.12.022}.

\bibitem{Blavatska09b}
Blavatska V., Janke W.,
Phys. Rev. E, 2009, \textbf{80}, 051805; \doi{10.1103/PhysRevE.80.051805}.

\bibitem{Blavatska08a}
Blavatska V., Janke W.,
Europhys. Lett., 2008, \textbf{82}, 66006; \doi{10.1209/0295-5075/82/66006}.

\bibitem{Blavatska08b}
Blavatska V., Janke W.,
Phys. Rev. Lett., 2008, \textbf{101}, 125701; \doi{10.1103/PhysRevLett.101.125701}.

\bibitem{Blavatska08c}
Blavatska V., Janke W.,
In: Proceedings of the International Conference ``Path
Integrals~--- New Trends and Perspectives'' (Dresden, 2007),
Janke W., Pelster A. (Eds.), World Scientific, Singapore, 2008, 585--588.

\bibitem{viktoria_jpa}
Blavatska V., Janke W.,
J. Phys. A: Math. Theor., 2009, \textbf{42}, 015001; \doi{10.1088/1751-8113/42/1/015001}.

\bibitem{Blavatska10b}
Blavatska V., Janke W.,
Physics Procedia, 2010, \textbf{3}, 1431; \doi{10.1016/j.phpro.2010.01.202}.

\bibitem{Fricke2013}
Fricke N., Janke W.,
Eur. Phys. J. Special Topics, 2013, \textbf{216}, 175; \doi{10.1140/epjst/e2013-01740-4}.

\bibitem{Rintoul94}
Rintoul M.D., Moon J., Nakanishi H., Phys. Rev. E, 1994, \textbf{49}, 2790; \doi{10.1103/PhysRevE.49.2790}.

\bibitem{Blavatska04}
Von Ferber C., Blavatska V., Folk R., Holovatch Yu.,
Phys. Rev. E, 2004, \textbf{70}, 035104(R); \\ \doi{10.1103/PhysRevE.70.035104}.

\bibitem{Janssen07}
Janssen H.-K., Stenull O., Phys. Rev. E, 2007, \textbf{75}, 020801(R); \doi{10.1103/PhysRevE.75.020801}.

\bibitem{Woo91}
Woo K.Y., Lee S.B., Phys. Rev. A, 1991, \textbf{44}, 999; \doi{10.1103/PhysRevA.44.999}.

\bibitem{Grassberger93}
Grassberger P., J. Phys. A: Math. Gen., 1993, \textbf{26}, 1023; \doi{10.1088/0305-4470/26/5/022}.

\bibitem{Lee96}
Lee S.B., J. Korean Phys. Soc., 1996, \textbf{29}, 1.

\bibitem{Havlin87}
Havlin S., Ben Abraham D., Adv. Phys., 1987, \textbf{36}, 695; \doi{10.1080/00018738700101072}.

\bibitem{viktoria_surface}
Blavatska V., Janke W.,
J. Chem. Phys., 2012, \textbf{136}, 104907; \doi{10.1063/1.3691102}.

\bibitem{viktoria_surface_athens}
Blavatska V., Janke W.,
Physics Procedia, 2012, \textbf{34}, 55; \doi{10.1016/j.phpro.2012.05.009}.

\bibitem{kumar-stretching}
Kumar S., Giri D., Phys. Rev. Lett., 2007, \textbf{98}, 048101; \doi{10.1103/PhysRevLett.98.048101}.

\bibitem{kumar-stretching2}
Kumar S., Jensen I., Jacobsen J.L., Guttmann A.J.,
Phys. Rev. Lett., 2007, \textbf{98}, 128101; \\ \doi{10.1103/PhysRevLett.98.128101}.


\bibitem{niklas-johannes}
Fricke N., Bock J., Janke W.,
diffusion-fundamentals.org, 2013, \textbf{20}, 111.


\end{thebibliography}

\ukrainianpart

\title{Полімери у невпорядкованих середовищах}

\author{В. Блавацька\refaddr{label1},
Н. Фріке\refaddr{label2}, В. Янке\refaddr{label2,label3}
}
\addresses{
\addr{label1} Інститут фізики конденсованих систем Національної академії наук України, 79011 Львів, Україна
\addr{label2} Інститут теоретичної фізики, Університет Лейпцига,
04009 Лейпциг, Німеччина
\addr{label3} Центр теоретичних досліджень (NTZ), Університет Лейпцига,
 04009 Лейпциг, Німеччина
}

\makeukrtitle

\begin{abstract}
\tolerance=3000%
Подаємо короткий огляд наших недавніх досліджень, метою яких є краще розуміння скейлінгової поведінки
полімерів у невпорядкованих середовищах.
Основна увага приділяється простій узагальненій моделі, в рамках якої полімери представляються
як (самовзаємодіючі) випадкові блукання із самоперетинами,  а невпорядковане середовище~--- як критичний перколяційний кластер.
Проаналізовано скейлінгову поведінку кількості можливих конформацій та усередненого просторового видовження полімера як
функції кількості мономерів, а також отримано значення відповідних критичних показників $\gamma$ та
$\nu$, застосовуючи два взаємодоповнюючі підходи: чисельні симуляції в рамках алгоритму зростаючого ланцюжка
(PERM) та точний перерахунок всіх можливих конформацій із використанням недавно розвиненої дуже ефективної методики.

\keywords  випадкові блукання, перколяційний кластер, комп'ютерні симуляції зростаючого ланцюжка (PERM), методи точного перерахунку
\end{abstract}
\end{document}